\documentclass[twocolumn,showpacs,preprintnumbers,amsmath,amssymb]{revtex4}

\begin{document}

\preprint{ANL-HEP-PR-06-37}

\title{Naturalness of the Vacuum Energy in Holographic Theories}

\author{Csaba Bal\'azs}
\affiliation{HEP Division, Argonne National Laboratory,
             9700 Cass Ave., Argonne, IL 60439, USA}
\email{balazs@hep.anl.gov}
\author{Istv\'an Szapudi}
\affiliation{Institute for Astronomy, University of Hawaii, 
             2680 Woodlawn Dr., Honolulu, HI 96822, USA} 
\email{szapudi@ifa.hawaii.edu}

\date{\today}

\begin{abstract}

Based on the cosmic holographic conjecture of Fischler and Susskind, we 
point out that the average energy density of the universe is bound from 
above by its entropy limit. Since Friedmann's equation saturates this 
relation, the measured value  of the cosmological energy density is 
completely natural
in the framework of holographic thermodynamics: vacuum energy 
density fills the available quantum degrees of freedom allowed by the 
holographic bound. This is in strong contrast with traditional quantum field 
theories where, since no similar bound applies, the natural value of the 
vacuum energy is expected to be 123 orders of magnitude higher than the 
holographic value. Based on our simple calculation, holographic 
thermodynamics, and consequently any future holographic 
quantum (gravity) theory, resolves the vacuum energy puzzle.

%

\end{abstract}


\maketitle

\section{Introduction}

The perceived accelerating expansion of the universe is the most mysterious 
puzzle of contemporary theoretical physics.  The standard interpretation of 
this acceleration requires the presence of a substance with negative 
pressure, i.e. vacuum energy or dark energy. The origin and nature of matter 
with such exotic equation of state is presently not understood.
Recent measurements of the temperature fluctuations in the cosmic microwave 
background \cite{Spergel:2003cb} and the Hubble diagram of supernovae 
\cite{Perlmutter:1997zf,Riess:1998cb} indicate that the present average 
cosmological vacuum energy density is close to the critical density
\begin{equation}
\rho_c = \frac{3}{8\pi} H_0^2 M_{P}^2,
\label{Eq:Friedmann0}
\end{equation}
where $H_0 = 1.24 \times 10^{-61} M_P$ is the Hubble constant and $M_P = 
1.22 \times 10^{19}$ GeV is the Planck mass \endnote{ Throughout this paper, 
we equate the Boltzmann and reduced Planck constants to the speed of light, 
as $k=\hbar=c=1$, and express dimensional quantities in $M_P$ or GeV units.}. 
Expressed in Planck units $\rho_c = 1.84 \times 10^{-123}$.  This 
astonishingly tiny energy density is much smaller than any fundamental 
physical scales so understanding its stability against quantum corrections 
is a serious problem \cite{Weinberg:1988cp}.  Perhaps the 
only known physical quantity that comes close to this number is the inverse 
entropy limit of the universe, which is estimated to be $10^{-123}$ in 
Planck units.

The minute size of the vacuum energy density poses at least three 
``cosmological constant'' problems:

$\bullet$ Why is the vacuum energy small?

$\bullet$ Why is the vacuum energy non-zero?

$\bullet$ Why are the vacuum and matter energy densities about the same today?

\noindent
There are various attempts to reconcile these vacuum energy problems with 
gravity and quantum physics (for recent reviews see for example 
\cite{Nobbenhuis:2004wn} or \cite{Copeland:2006wr} and references therein). 
A possible avenue to solve the first problem is the utilization of a 
fundamental symmetry which readily renders the vacuum energy zero.  To 
explain the size of the vacuum energy some landscape or multiverse scenarios 
derive it from anthropic considerations \cite{Weinberg:1987dv,Susskind:2003kw}.
Another class of proposed solutions introduces a scalar field that couples to 
gravity such that the quantum fluctuations are cancelled when the scalar 
field reaches its equilibrium value.  Recently, much attention has been 
focused on quintessence models \cite{Caldwell:1997ii}. 

In this work, we are only concerned with the first problem. What explains 
the naively unnatural smallness of the critical density? Why does quantum 
field theory overestimate the vacuum energy density by 123 orders of 
magnitude? What protects the vacuum energy from becoming the order of the 
Planck scale? 
We argue that in holographic theories \cite{'tHooft:1993gx} the entropy 
bound sets an upper limit on the average energy density. Based on this, we 
derive the energy density of the universe from the Fischler-Susskind entropy 
limit \cite{Fischler:1998st}. We show that the energy density of the 
universe is bound by the inverse of its entropy when the holographic bound
is saturated.  Thus in the 
framework of holographic theories the solution of the first vacuum energy 
problem is profoundly simple: the amount of quantum fluctuations (and matter 
energy) is capped by the entropy limit of the universe. 

General relativity is the prime example of a holographic theory 
\cite{Jacobson:1995ab}. As a consequence of this, Friedmann's equation 
governing the evolution of the universe is {\it known to saturate the 
holographic bound} \cite{Bousso:1999xy} for a wide range of cosmological 
parameters \cite{Bak:1999hd}. We show that holography is the primary reason 
that the dynamic argument based on Friedmann's equation yields the smallness
of the total energy density and thereby the vacuum energy. On the 
contrary, quantum field theories in their present form are not holographic. 
This is why they overestimate the vacuum energy by an enormous magnitude, 
related to the ratio of the horizon area to the Planck length squared. As 
quantum fluctuations fill the available quantum degrees of freedom, the 
observed increase of the vacuum energy is natural in an expanding universe.

\section{The vacuum energy problem}

To quantify the vacuum energy problem, we seek the average 
energy density, $\rho = E/V$, of a homogeneous, spherical system 
occupying a volume $V = 4 \pi R^3/3$. We assume that the total 
angular momentum and charge of the system are negligible.
In  any extensive theory, such as quantum field theories,
we can stack (real or virtual) energy quanta of Planck size 
$M_P$ at Planck length $L_P = 1/M_P$ distances from each other to
fill a volume
\cite{Bousso:2002ju}, such that the total energy is $E \sim (V/L_P^3)M_P$.
This idea can be formalized as a Planck scale ultraviolet regularization,
and leads to an enormous energy density
\begin{eqnarray}
\rho \sim M_{P}^{4}.
\end{eqnarray}

This result is typically contrasted with the critical density 
(\ref{Eq:Friedmann0}) which is, indeed, about 123 orders of magnitude 
smaller than the quantum field theory estimate. Friedmann's equation 
(\ref{Eq:Friedmann0}), relating the energy density to the Hubble constant, 
follows from general relativity. This begs the question:
why does general relativity relate the energy density to a 
fundamental parameter of the universe `correctly', while quantum field theory 
grossly fails?

\section{The vacuum energy: no problem}

We argue that the smallness of the observed cosmological vacuum energy 
density is a simple and natural consequence of two {\it known} facts:

$\bullet$ the universe is a holographic system, i.e. its entropy is limited 
by a holographic entropy bound, and

$\bullet$ the universe is large (compared to the Planck length).

\noindent
To illustrate this, we consider the following example.
While trying to pile Planck energy quantum oscillators at Planck distances, 
it is easy to realize that gravity will strongly limit the total energy in a 
given volume. Since the radius of a Planck mass Schwarzschild black hole is 
$L_P M_P^2/2$, when placing two of them close enough to each other they form 
a black hole with radius $L_P M_P^2$. Any extra energy in the $4 \pi 
L_P^3/3$ volume is gravitationally unstable. Thus, the energy of the 
gravitationally stable configurations in a given volume is limited by the 
{\it radius}, rather than the volume. 

When assuming that an $R^3$ volume can be filled by $M_P$ quanta at $L_P$ 
distances, we overestimate the vacuum energy density roughly by a factor of 
$r \sim (2 R/L_P)^3/(2 R/L_P)$. In the case of a black hole with the present 
size of the universe this factor is
\begin{eqnarray}
r \sim \frac{4 M_P^2}{H_0^2} \sim 3 \times {10}^{122},
\end{eqnarray}
where we equated the radius with the apparent  horizon
of the universe.

While it is obvious for a Schwarzschild black hole, we argue that it is a 
generic property of holographic systems that their total energy scales with 
linear size as a straightforward consequence of holography. More precisely, 
a system that saturates the holographic bound also satisfies the 
Schwarzschild condition, i.e. its maximal mass is the half of its radius in 
Planck units. To show this, we consider a system with total energy $E$. 
The system is to saturate the holographic entropy bound, so we are looking for 
its maximum entropy. Since the entropy of the system is bound from above by 
the entropy of a black hole, its maximal entropy is
\begin{eqnarray}
S = \pi {R^2} M_P^2.
\end{eqnarray}
If the system saturates Bekenstein's entropy bound
\begin{eqnarray}
S \leq 2 \pi E R,
\end{eqnarray}
then
\begin{eqnarray}
E = \frac{R }{2} M_P^2,
\end{eqnarray}
that is the Schwarzschild condition is satisfied by the system.

Let us calculate the average energy density 
$\rho$ of a homogeneous, spherical system that saturates the holographic 
entropy bound. Since the mass of any gravitationally stable holographic 
system is bound from above by the Schwarzschild condition, the maximal 
energy density of the system
\begin{eqnarray}
\rho = \frac{3}{8 \pi  R^2} M_P^2
\end{eqnarray}
is the inverse function of its size. This implies that a small system with 
\(R={L_P}\) has an upper limit \(\rho \) \(=\) 3\(M_P^4\)/8\(\pi \) on its 
energy density, as naively expected in field theory. On the other hand, a 
larger system will necessarily have a smaller energy density. This is the 
simple consequence of holography: energy scales with linear size, so the 
energy density decreases with the area. This scaling is
consistent
with the smallness of the measured cosmological energy density: a system 
the size of the universe has a rather stringent upper limit on its energy 
density. The holographic principle reconciles the field theory Planck scale 
cutoff with the smallness of the cosmological vacuum energy density in a 
remarkably simple manner. 

To show that the above limit on the energy density will also include the 
energy density of the vacuum, consider a Schwarzschild-de Sitter black hole 
with mass $M$ and cosmological constant $\Lambda$. The energy contained 
inside its horizons is
\begin{eqnarray}
E = M+\Lambda \frac{{R^3}}{3} = \frac{R}{2} M_P^2.
\end{eqnarray}
We introduce the matter and vacuum energy densities as $\rho_M = M/V$ and 
$\rho_V = \Lambda/(4\pi)$, which leads to
\begin{eqnarray}
\rho = \rho_M + \rho_V = \frac{3}{8 \pi {R^2}} M_P^2.
\end{eqnarray}
Since any gravitating system with the same radius and higher energy content 
is gravitationally unstable, we consider the right hand side of this 
equation as an upper limit on the average total energy density of a physical 
system in de Sitter space.  Our result is strengthened by the observation 
that quantum corrections in holographic theories follow exactly the same 
$1/R^2$ scaling \cite{Cohen:1998zx,Thomas:2000km,Hsu:2004ri}.

These results suggest that in gravitating quantum systems the vacuum 
contribution to the total energy density is strongly limited by gravity. 
Assuming that this holographic limit holds for the entire universe,
the worst fine tuning problem in the history of physics appears
to be naturally eliminated.

\section{Holographic limit on the energy density of the universe}

While the previous results are intriguing, they are not obviously applicable 
to the universe: the corresponding Friedman-Robertson-Walker metric of the 
universe is different from that of a Schwarzschild (-de Sitter) black hole. 
Moreover, an argument based on gravity would not be independent of the 
dynamic argument based on Friedmann's equation. In this section, we show is 
that the holographic conjecture combined with thermodynamics sets the {\it 
same} limit on the vacuum energy that follows from the Einstein-Friedmann
equations. Furthermore, the 
present value saturates this limit, without any fine tuning. 

We argue that the smallness of the energy density for large black holes 
and the universe are all the consequence of the general holographic entropy 
limit. To clarify this, we sketch the derivation of the cosmological energy 
density from the holographic conjecture without relying on Einstein's 
equation.
We start with the Fischler-Susskind cosmic holographic conjecture 
\cite{Fischler:1998st}: the entropy of the universe is limited by its 
`surface' measured in Planck units
\begin{eqnarray}
S \leq \frac{A}{4} M_P^2,
\label{Eq:FS}
\end{eqnarray}
where the surface area $A = 4\pi R^2$, motivated by causality, is defined in 
terms of the apparent (Hubble) horizon
\begin{eqnarray}
R = \frac{1}{\sqrt{H^2+k/a^2}},
\end{eqnarray}
with curvature and scale factors $k$ and $a$, respectively.
According to equation (\ref{Eq:FS}) the average entropy density is limited by
\begin{eqnarray}
\frac{S}{V}\leq \frac{\pi R^2}{V} M_P^2.
\label{Eq:FSSV}
\end{eqnarray}

The first law of holographic thermodynamics relates the entropy and 
energy of a  holographic system,
\cite{Hawking:1974sw,Jacobson:1995ab,Cai:2005ra}
\begin{eqnarray}
E = T S,
\label{Eq:1stLaw}
\end{eqnarray}
where we assumed that the universe has no net angular momentum or 
electric charge.
From equations (\ref{Eq:FSSV}) and (\ref{Eq:1stLaw}), we obtain
\begin{eqnarray}
\rho \leq \frac{3 T}{4 R} M_P^2.
\end{eqnarray}
This is the holographic limit on the energy density of the universe
in terms of the temperature and horizon.

The natural value of the temperature in an adiabatically expanding system is 
$T \simeq 1/R$, perhaps with a constant of proportionality of order unity. 
We fix the constant by using the Gibbons-Hawking temperature 
\cite{Gibbons:1977mu} of the horizon of the universe 
\cite{Padmanabhan:2002ha,Padmanabhan:2004tz}
\begin{eqnarray}
T = \frac{1}{2\pi R},
\label{Eq:GH}
\end{eqnarray}
which sets the ${\cal O}(1)$ constant to be $1/2\pi$. 
In this context, naive usage of the measured temperature of the cosmic 
microwave 
background (or neutrinos) would constitute a fine tuning  of
an order $10^{30}$. Moreover, this substitution would yield a much weaker limit 
than the  natural value (although it would be still $10^{90}$ times 
tighter than the traditional field  theory estimate). 
Using equation (\ref{Eq:GH}), we obtain
\begin{eqnarray}
\rho \leq \frac{3}{8 \pi R^2} M_P^2 = \frac{3}{2 A} M_P^2,
\label{Eq:rhoA}
\end{eqnarray}
that is the energy density of the universe is bound by the inverse area of 
its horizon. 

From general relativity, we know that the universe saturates the holographic 
bound \cite{Bousso:1999xy}. This is a natural expectation from 
quantum physics: if matter contributes less, quantum fluctuations will 
maximize the entropy of the universe.  Then, using the equal sign, 
equation (\ref{Eq:rhoA}) yields Friedmann's equation
\begin{eqnarray}
\rho = \frac{3}{8 S} M_P^4,
\label{Eq:FriedmannS}
\end{eqnarray}
stating that the energy density of the universe is bound by its entropy limit
\begin{eqnarray}
S = \frac{\pi}{H^2 + k/a^2}.
\end{eqnarray}
The derivation of Friedmann's equation from thermodynamics 
and the cosmological holographic limit can be performed more rigorously for 
various cosmological models with results identical to ours 
\cite{Cai:2005ra,Akbar:2006er}.

The presence and otherwise mysterious increase of the vacuum energy density 
is also consistent with the holographic picture. Due to the expansion of the 
universe the total energy density dilutes as $1/R^3$, while the upper limit 
set by gravity decreases only as $1/R^2$. The new quantum degrees of freedom 
created by the increase of the horizon area are continuously filled by 
vacuum energy.

Connecting the energy density and the number of quantum states of a 
gravitating system, relation (\ref{Eq:FriedmannS}) can be reinterpreted as an
equation of quantum gravity, perhaps signaling the importance of the 
holographic principle for the quantized description of gravitation.  
Details of such a future theory are unknown at this time, but it
is most likely to obey the holographic conjecture.
The covariant form of 
(\ref{Eq:FriedmannS}), possibly a 
modification of Einstein's equations, might be a 
step toward an effective field theory of quantum gravity. We also speculate 
that the appearance of the inverse entropy (or inverse mass squared) might 
also signal a (self-)duality of quantum gravity.

Finally, we make it explicit where the apparent $10^{123}$ factor comes from. 
It is simply the ratio \(r\) of the holographic entropy limit of the 
universe to the entropy of a Plank mass and length size Schwarzschild black 
hole: 
\begin{eqnarray}
r = M_P^4 \left(\frac{3{M_P^2}}{8\pi R^2}\right)^{-1}
  =\frac{8 \pi}{3} \frac{R^2}{L_P^2}
  =\frac{8 \pi}{3} \frac{S}{S_P}
  \propto \log(N),
\nonumber
\end{eqnarray}
where $N$ is the available (micro) quantum states of the universe. 
Indeed, substituting numerical values, we obtain
\begin{eqnarray}
r = 5.44\times 10^{122}.
\end{eqnarray}

\section{Conclusions}

Holographic thermodynamics imposes an upper limit on the energy 
density of the universe by its entropy bound. Friedmann's equation
saturates this limit, therefore the inferred value of the vacuum
energy from cosmological observations of the expansion rate is
naturally consistent with the holographic context. 
In fact, this saturation has an intuitive meaning 
of the vacuum energy filling up all available quantum degrees of freedom 
as the universe expands. While holography
is a conjecture, it provides general 
framework which future quantum gravity theories are likely to fill with 
microphysical content. Thus our simple thermodynamic estimates are expected to
remain valid in quantum gravity, although
the technology of the calculation is likely to be more complex.

The simplicity of the holographic limit on the vacuum energy is contrasted 
with the mysterious $10^{123}$ discrepancy between the quantum field theory 
estimate and the measured cosmological energy density. 
This is yet another argument strengthening the case
for the holographic conjecture, and motivating the search for
holographic theories. Since holography has its roots in quantum theory,
our considerations suggest that Friedmann's equation, although
originates from a classical field theory, has a corresponding 
quantum interpretation connecting the energy density with the 
available quantum states of the universe.

\bigskip
\begin{acknowledgments}

We thank Josh Frieman, Nick Kaiser and Robert Wald for useful suggestions. 
IS was supported by NASA through AISR NAG5-11996, and ATP NASA NAG5-12101 as 
well as by NSF grants AST02-06243, AST-0434413 and ITR 1120201-128440. 
Research at the HEP Division of ANL is supported in part by the US DOE, 
Division of HEP, Contract W-31-109-ENG-38. CB also thanks the Aspen Center 
for Physics for its hospitality and financial support.

\end{acknowledgments}

\bibliography{holography}

\end{document}